# A New Result on the Random Residue Sequence Algorithm

Vamsi Sashank Kotagiri
Oklahoma State University, Stillwater

**Abstract**
Random residue sequences (RR) may be used in many random number applications including those related to multiple access in communications. This paper investigates variations on an algorithm to generate RR sequences that was proposed earlier by the author. This makes it possible to obtain many more random sequences than was possible to do by the previous algorithm. Experimental results are presented on a variety of sequences of length 16 and 24. To obtain a variety of RR sequences of a specific length can have obvious applications in cryptography.

*Keywords: Random residue sequences, random numbers, random access communications, cryptography*

**Introduction**
The motivation to study random residue sequences (RR) is the use of residue sequences in place of binary random sequences in a variety of cryptographic and communications applications [1]-[4]. The starting point in this study is the discrete Hilbert transforms [5]-[7] which make it possible to obtain a sequence that is orthogonal in phase. These sequences are used in statistical and complex signal analysis [8],[9]. Recently, this theory has been generalized to the number theoretic Hilbert transform (NHT) [10]-[15].

When the NHT matrix is multiplied with its transpose (which is its inverse) for a particular prime modulus n it computes all correlations on the block which is what makes it possible to generate random residue sequences. These sequences may be used in place of scrambling [16],[17] and d-sequences [8]-[21].

RR sequences may be used as carriers in wireless communication systems [11]. For a data sequence *a(n)* of N points the autocorrelation function *C(k)* is represented by

$$C_a(k) = \frac{1}{N} \sum_{j=1}^{N} a(j)a(j+k) \qquad (1)$$

For a noise sequence, the autocorrelation function $C_a(k) = E(a(i)a(i+k))$ is two-valued, with value of 1 for k=0 and a value approaching zero for k≠0 for a zero-mean random variable. The autocorrelation of RR sequences is also two valued, that is, it is non-zero for k=0 and multiples of N, and zero for all other values. This property makes the application of RR sequences to multiple access systems obvious.

An algorithm was proposed in [12] for generating the ideal random residue sequence. For obtaining a RR sequence we select a prime number as the first element and the remaining elements as powers of 2 in a number theoretic Hilbert transform matrix with the alternating zeroes suppressed. This sequence is given as input to the algorithm and as a result we obtain both an RR sequence together with its prime modulus n.

$$C_a(k) = \frac{1}{N} \sum_{j=1}^{N} a(j)a(j+k) \mod n \tag{2}$$

The algorithm in (12) helped us find many RR sequences but as the length of the sequence desired increases, the algorithm fails in many situations. This leads to the question: What are the parameters under which the algorithm works efficiently?

In this paper, we present our findings that relate to considering variants of the previous algorithm that generate successful RR sequences.

**Obtaining the Modulus for the RR Sequence**
The process of obtaining the modulus for the RR sequence is as follows:

1. Pick the row in the circulant matrix using our generation algorithm
2. Compute the autocorrelation function given in equation (1)
3. Find the g.c.d. of the non-zero autocorrelation values in $C(k)$ for non-zero values of $k$.
4. If the g.c.d. is 1, RR sequence does not exist.
5. If the g.c.d. is a prime number then that is the modulus.
6. If the g.c.d. is a composite number, then consider one of its prime factors as the prime modulus.

The prime modulus obtained would be efficient if it is the same size as the length of the sequence or smaller. Its optimal value will be determined by the application in mind.

In equation (2), when n=2, we obtain random sequence results for binary sequences. A general orthogonal binary sequence will not exist if N=odd as well as when N=even.

A question of some interest is the relationship between N and good values of the prime modulus n, especially for small values of the latter.

**The Variant Algorithm**
The variation of the previous algorithm is primarily in the use of prime other than 2, together with its powers, in the choice of the first row of the number theoretic Hilbert transform. We perform an experiment where we run through a set of starting prime numbers; and we have chosen prime numbers up to 100.

Table 1. Moduli obtained using different starting primes in the generation algorithm for sequences of length 16

| No | Starting prime number | Prime modulus obtained |
|----|----------------------|------------------------|
| 1  | 2                    | 331                    |
| 2  | 3                    | 3121                   |
| 3  | 5                    | 7283                   |
| 4  | 7                    | 21851                  |
| 5  | 11                   | 47                     |
| 6  | 13                   | 1987                   |
| 7  | 17                   | 347                    |
| 8  | 19                   | 21863                  |
| 9  | 23                   | 197                    |
| 10 | 29                   | 317                    |
| 11 | 31                   | 7                      |
| 12 | 37                   | 21881                  |
| 13 | 41                   | 1459                   |
| 14 | 43                   | 509                    |
| 15 | 47                   | 7297                   |
| 16 | 53                   | 811                    |
| 17 | 59                   | 149                    |
| 18 | 61                   | 337                    |
| 19 | 67                   | 21911                  |
| 20 | 71                   | 487                    |
| 21 | 73                   | 101                    |
| 22 | 83                   | 7309                   |
| 23 | 89                   | 2437                   |

The results are given in Table 1.

The length of the sequence for the above table is 16 long random residue sequence. When the above values are plotted on a graph, the graph will be as follows where the X-axis represents starting prime chosen and Y-axis represents the obtained prime modulus.

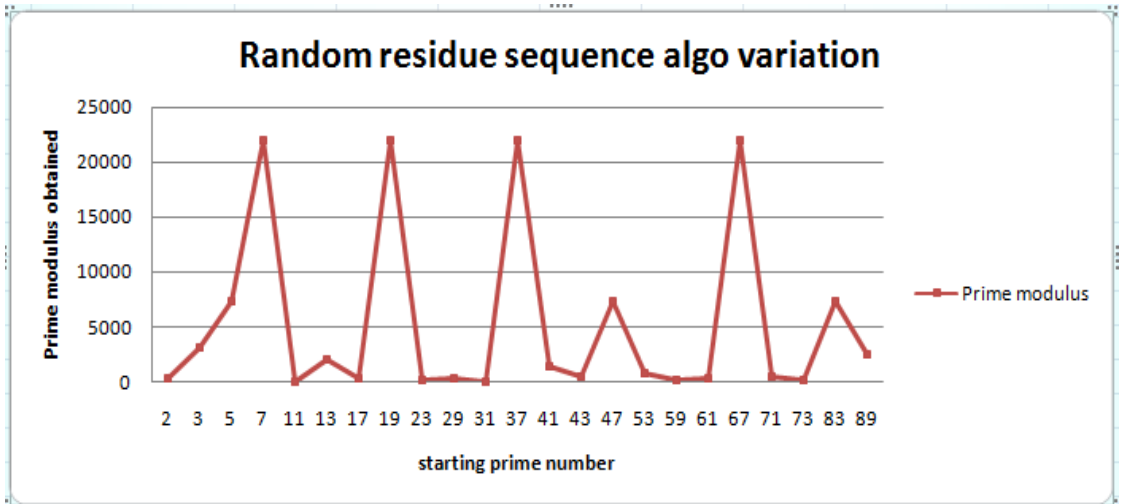
Figure 1. Modulus versus starting prime number

Table 2 presents similar results for sequences of length 24, again for primes less than 100.

Table 2. Moduli obtained using different starting primes in the generation algorithm for sequences of length 24

| No | Starting prime number | Prime modulus obtained |
|---|---|---|
| 1 | 2 | 113 |
| 2 | 3 | 2521 |
| 3 | 5 | 32779 |
| 4 | 7 | 83 |
| 5 | 11 | 32797 |
| 6 | 13 | 32803 |
| 7 | 17 | 6563 |
| 8 | 19 | 1427 |
| 9 | 23 | 32833 |
| 10 | 29 | 19 |
| 11 | 31 | 103 |
| 12 | 37 | 263 |
| 13 | 41 | 32887 |
| 14 | 43 | 127 |
| 15 | 47 | 6581 |
| 16 | 53 | 73 |
| 17 | 59 | 32941 |
| 18 | 61 | 701 |
| 19 | 67 | 347 |
| 20 | 71 | 673 |
| 21 | 73 | 32983 |
| 22 | 83 | 33013 |
| 23 | 89 | 67 |

There is an interesting pattern to the value of the prime modulus obtained as the starting prime number is changed. To have a small modulus is clearly desirable from an implementation point of view and therefore the conditions under which small moduli are obtained is a problem of obvious interest.

**Conclusions**

This paper presented the results obtained by starting with prime numbers other than 2 and finding what subsequent prime moduli are obtained as result. This research opens up many questions for further work:

- Is the relationship between the starting prime and the prime modulus obtained using the algorithm to generate RR of a regular nature that can be exploited in quick generation of RR sequences?
- How does the algorithm work as the value of n is increased?
- Can very large n RR sequences be generated in an efficient manner?
- What are the general properties of RR sequences?

The answer to these questions will be related to the properties of constrained circulant matrices.